\documentclass[prb,aps,twocolumn,superscriptaddress,showpacs,longbibliography]{revtex4-2}
\usepackage{natbib}
\usepackage{graphicx}
\usepackage{amssymb}
\usepackage{wrapfig}
\usepackage{epstopdf}
\usepackage{amsmath}
\usepackage{xspace} 
\usepackage{color}
\newcommand{\nham}{\textcolor{black}}
\newcommand{\nhamc}{\textcolor{black}}

\begin{document}
\title{Topological spin-up triplet excitonic condensation in two-dimensional electron-hole systems}
\author{Van-Nham Phan}
\thanks{{Corresponding author: phanvannham@duytan.edu.vn}}
\affiliation{Institute of Research and Development, Duy Tan University, 3 Quang Trung, Danang  550000, Vietnam}
\affiliation{Faculty of Natural Sciences, Duy Tan University, 3 Quang Trung, Danang 550000, Vietnam}
\begin{abstract}
We investigate topological spin-up triplet excitonic condensation and its competition with other stabilities in a two-dimensional interacting electron-hole system taking into account Rashba spin-orbit coupling and external magnetic fields. Using an unrestricted Hartree-Fock approach, we self-consistently evaluate spin-selective excitonic condensate order parameters and the Chern number. The ground state phase diagram in the dependence on magnetic field and Coulomb interaction shows a spin-up triplet excitonic condensate (EC) with a nonzero Chern number, emerging uniquely away from the topologically trivial singlet and spin-down triplet EC regions. Strong spin-polarized triplet excitonic fluctuations preceding the condensation are further revealed through the signatures of the dynamical excitonic susceptibility spectra. Our results establish a class of topological quantum phases driven by excitonic coherence and suggest a realistic pathway to its realization in a \nham{distorted Janus monolayer of transition metal dichalcogenides or some twisted van der Waals heterostructures}.
\end{abstract}
\date{\today}
\maketitle

{\it Introduction.}---The discovery of topological phases of matter has revolutionized our understanding of condensed matter physics, unveiling a rich landscape of quantum states characterized by robust boundary phenomena and nontrivial topological invariants~\cite{RMP.82.3045,RMP.83.1057,NRM.7.196}. The quantum Hall effect in two-dimensional (2D) systems exemplifies this paradigm, where a perpendicular magnetic field quantizes electronic states into Landau levels (LLs), breaking time reversal symmetry (TRS) and yielding a Hall conductance proportional to an integer Chern number that counts chiral edge states~\cite{RMP.82.3045,RMP.83.1057,NRM.7.196,RMP.59.781,RMP.71.S298,NRM.2.397,RMP.75.1101}. These nontrivial topological phases are fundamentally noninteracting and arise purely from the band structure. In the meanwhile, the strong electron-hole interaction plays significant roles in 2D systems, particularly in semimetal (SM) and semiconducting (SC) platforms where excitons, bound states of electrons and holes, can condense into macroscopic quantum states under specific conditions~\cite{Mo61,NC82,KRV17,NatCommu.12.1969}. Such excitonic condensation (EC), driven by the Coulomb interactions, has been extensively explored in double-layer quantum Hall systems, revealing phenomena such as coherent interlayer tunneling and quantized Hall drag resistance as hallmarks of the condensate, underlining the interplay of topology and interactions~\cite{Nat.635.301,RMP.88.025003,PRB.107.115117,PRB.103.L201115,PRL.104.016801,PRX.13.031023,PRL.134.046601}. 

Transition metal dichalcogenide (TMD) \nham{monolayers or some twisted van der Waals heterostructures}, with their tunable \nham{narrow-band} structures, strong spin-orbit coupling (SOC), \nham{large electron-hole interaction} and sensitivity to external fields, have recently emerged as promising platforms for exploring such phenomena~\cite{PRB.84.153402,PRL.108.196802,NN.15.367,NC.9.1427,PRL.131.046402,PRB.111.155139,PRB.103.195114,PhysE.134.114934}. \nham{In these systems}, a magnetic field induces LL quantization, while SOC introduces spin-split bands that can enhance the topological character of the system~\cite{PRL.131.046402,CP.1.40,AP.435.168561,PRB.110.115114}. When electron-hole pairs form excitons with a triplet spin configuration, the resulting condensate may exhibit topological properties distinct from the singlet-dominated states typically observed in quantum spin Hall insulators (QSHIs) or bilayer quantum Hall systems~\cite{PRB.109.075167,PRL.124.166401,PRB.103.L201115,NC.10.210}. Spin-triplet EC states are particularly promising for quantum technologies due to their distinct advantages over spin-singlet counterparts. Unlike singlet excitons, which dominate in conventional SC materials due to $s$-wave pairing, triplet excitons support spin superfluidity, enabling dissipationless spin transport~\cite{PRB.84.214501}. Their spin-polarized nature and longer radiative lifetimes, arising from optical selection rules, enhance coherence times, making them ideal for quantum information processing and storage~\cite{NC.14.1180,JPCC.123.18665,PRX.5.011009}. Additionally, triplet excitons can couple to topological band structures, yielding nontrivial invariants such as Chern numbers or $\mathbb{Z}_2$ indices, which facilitate robust quantum states resistant to decoherence~\cite{PRL.121.126601,PNAS.121.e2401644121,PRB.102.205124,PRB.105.155112}. These properties position the spin-triplet ECs as candidates for fault-tolerant quantum computing and spin-based quantum sensors, surpassing the limitations of singlet ones, which lack spin degrees of freedom and topological protection~\cite{PNAS.121.e2401644121,Sci.365.684}. 

\nham{Recent theoretical advances have uncovered diverse microscopic origins of topological spin-triplet ECs. Indeed, in flat-band kagome lattices, highly localized electron-hole wavefunctions and suppressed screening were shown to greatly enhance binding energies and stabilize a triplet EC with spin superfluidity~\cite{PRL.126.196403}. On topological-insulator Dirac surfaces, Wang {\it et al.} predicted a chiral $p+ip$ spin-triplet EC characterized by a fractional Chern number and parity anomaly~\cite{NC.10.210} while Blason and Fabrizio demonstrated that in the Bernevig-Hughes-Zhang model, Coulomb interactions can soften and condense $S^z=\pm 1$ spin-triplet excitons with finite Chern numbers, leading to a magnetoelectric excitonic phase~\cite{PRB.102.035146}. Building on these predictions, a recent high-field transport study has provided compelling evidence for a spin-triplet EC in the three-dimensional topological material HfTe$_5$, marked by a field-induced gap opening and vanishing Hall response in the ultra-quantum limit~\cite{PRL.135.046601}. Despite this progress, the microscopic role of magnetic fields and SOC in stabilizing topological spin-triplet ECs especially in 2D electron-hole systems remains largely unexplored. Clarifying this mechanism is crucial, as it could enable magnetically tunable, spin-superfluid excitonic phases with robust coherence and topological protection.}

\nham{Motivated by the pronounced Rashba SOC arising from the intrinsically broken inversion symmetry and strong electron-hole Coulomb interaction in Janus 1$T$-phase TMD monolayers, such as strained 1$T$-WSSe~\cite{PRB.103.195114,PRB.109.125108}, we investigate the topological spin-triplet EC state in a 2D interacting electron-hole system subject to an external magnetic field with incorporating the intraband Rashba SOCs}. Using the unrestricted Hartree-Fock approximation (UHFA), a ground-state phase diagram delineating the stability of EC states with all possible electron and hole spin configurations is constructed. \nham{At zero temperature, UHFA has been specified as an applicable method even at the strongly correlated limit~\cite{Georges06,Cz99,Fa08,PRB.95.045101}. Such a mean-field treatment has been widely used in prior studies linking SOC-induced band inversion and topological properties in the excitonic systems~\cite{PRL.98.166405,PRL.103.086404,PRB.82.195324,PRB.84.155447,PRB.109.075167}.}  Excitonic susceptibility functions \nham{evaluated in the random phase approximation (RPA)} further highlight the enhanced coherence of the triplet state near, \nham{but outside, the ECs}.  These findings elucidate the conditions for topological spin-up triplet ECs and pave the way for their experimental realization in magnetically tunable 2D platforms for advanced quantum technologies. \nham{In our situation, the Rashba SOC terms are incorporated in the UHFA, capturing the essential SOC-induced symmetry breaking, and are also explicitly included in the particle-hole matrix elements of the RPA kernel derived from the original many-body Hamiltonian. Consequently, SOC-induced $U(1)$-breaking processes and the resulting spin mixing are fully accounted for, making the combined methodology both quantitatively reliable and physically transparent.}.

{\it {Model and method}.}---\nham{The many-particle Hamiltonian capturing the essential physics of the band structure, SOC, external magnetic field, and electron-hole Coulomb interaction of a 2D lattice electron-hole system can be separated into a noninteracting part $\mathcal{H}_0$ and an interaction one $\mathcal{H}_{\text{int}}$~\cite{PNAS.121.e2401644121,PRL.121.126601},} such that
\begin{align}
\mathcal{H}_0&= \sum_{{\bf k}\alpha\sigma}\varepsilon^\alpha_{\bf k}\alpha^\dag_{{\bf k}\sigma}\alpha^{}_{{\bf k}\sigma}-\sum_{{\bf k}\alpha\sigma\sigma'}\mathcal{L}^\alpha_{\sigma\sigma'}({\bf k})\alpha^\dag_{{\bf k}\sigma}\alpha^{}_{{\bf k}\sigma'},
\label{1}
\end{align}
where, $\alpha_{\mathbf{k}\sigma}^\dagger$ and $\alpha_{\mathbf{k}\sigma}^{}$ are creation and annihilation operators for conduction ($\alpha=c$) and valence ($\alpha=f$) electrons with momentum $\mathbf{k}$ and spin $\sigma$, respectively. Note here that the valence electron operators can be transffered to that of valence holes by the standard electron-hole transformation. The first term in Eq.~\eqref{1} describes the kinetic energy of the electron system and in a 2D lattice with a tight-binding approximation, the dispersion relations read $ \varepsilon^{\alpha}_{\bf k}=-2t^{\alpha}(\cos k_x+\cos k_y)+E^\alpha-\mu$, with $E^\alpha$ is the onsite energy and $\mu$ is the chemical potential. In the second term, $\mathcal{L}^\alpha_{\sigma\sigma'}({\bf k})=\mu_BH\sigma^z_{\sigma\sigma'}+2\lambda^\alpha{\bf {L}}_0({\bf k}){\boldsymbol \sigma}_{\sigma\sigma'}$, where $H$ is the magnetic field, $\mu_B$ is the Bohr magneton, and $\lambda^\alpha$ is the strength of the intraband Rashba-SOC with ${\bf L}_0({\bf k})= (\sin k_y, -\sin k_x)$ and $\boldsymbol{\sigma}_{\sigma\sigma'}$ being the Pauli matrix vectors. \nhamc{In the $\mathcal{H}_0$ expression above, we have omitted the interband SOC components and orbital coupling. That simplification is applicable to the $1T$- and Janus TMDs. Indeed, in these structures, the SOC that arises from Rashba-type acting within the individual conduction and valence bands is dominant due to broken out-of-plane mirror symmetry~\cite{PRB.103.195114,PRB.97.235404}. The interband SOC components are typically symmetry-suppressed and much weaker. In the meanwhile, the combination of large effective masses, relatively flat band curvatures, and strong SOC ensures that the Zeeman interaction dominates the magnetic response, providing the principal source of band splitting and spin polarization relevant to excitonic coherence~\cite{PRB.84.153402,PRB.97.081109R,arXiv.2404.15134v1}. A full treatment will be addressed in future studies.}

\nhamc{In 2D TMD materials, the relevant low-energy states are dominated by transition metal $d$-orbitals localized on the same lattice site, making the local Coulomb interaction between conduction and valence electrons to become dominant~\cite{PRB.88.085440,iSci.26.106681,PRB.109.125108}. The on-site interaction thus captures the leading part of the excitonic binding. Including longer-range Coulomb components would primarily refine the quantitative details and such extensions will be explored in future work. Intraband Coulomb repulsion can be also incorporated but in the UHFA mainly renormalizes the quasiparticle dispersion without changing the qualitative nature of the condensate.} For this reason, we model the interaction Hamiltonian in the following form
\begin{equation}
\mathcal{H}_{\text{int}} = \frac{U}{N}\sum_{\mathbf{k}\mathbf{k}'\mathbf{q}, \sigma\sigma'}c_{\mathbf{k}\sigma}^\dagger c^{}_{\mathbf{k}+\mathbf{q},\sigma} f_{\mathbf{k}'\sigma'}^\dagger f^{}_{\mathbf{k}'-\mathbf{q}, \sigma'}. 
\end{equation}

To account for the ECs addressed in the total Hamiltonian, we employ the unrestricted Hartree-Fock (UHF) approximation, which allows for examining  the spin-dependence of the EC order parameters. In the approach, $\mathcal{H}_{\text{int}}$ can be rewritten in a form as $\mathcal{H}_{\text{int}}^{\text{UHF}} = U\sum_{\mathbf{k}\sigma} (n^cf^\dagger_{\mathbf{k}\sigma}f^{}_{\mathbf{k}\sigma}+n^fc^\dagger_{\mathbf{k}\sigma}c^{}_{\mathbf{k}\sigma})+\sum_{\mathbf{k}\sigma\sigma'}(\mathcal{D}_{\sigma\sigma'}c^\dagger_{\mathbf{k}\sigma} f^{}_{\mathbf{k}\sigma'}+ \text{H.c.})$, where $n^\alpha=\sum_{{\bf k}\sigma}\langle \alpha^\dag_{{\bf k}\sigma}\alpha^{}_{{\bf k}\sigma}\rangle/N$ is the conduction or valence electron density and $\mathcal{D}_{\sigma\sigma'}=-U\sum_{\mathbf{k}}\langle f^{\dagger}_{\mathbf{k}\sigma'} c^{}_{\mathbf{k}\sigma}\rangle/N$ represents the coherent pairing of an electron with spin $\sigma$ and a valence electron with spin $\sigma'$ or a valence hole with spin $-\sigma'$~\cite{NC.10.210}. $N$ here is the number of lattice sites. In the basis defined as $\Psi_{\mathbf{k}} = (c_{\mathbf{k}\uparrow}, c_{\mathbf{k}\downarrow}, f_{\mathbf{k}\uparrow}, f_{\mathbf{k}\downarrow})^T$, the full Hamiltonian $\mathcal{H}$ can be written in a $4\times 4$ Bogoliubov-de Gennes (BdG) matrix
\begin{align}
\mathcal{H}^\text{BdG}&(\mathbf{k})=\nonumber\\
&\begin{pmatrix} 
\bar{\varepsilon}^c_{\bf k}-\mu_BH & -2\lambda^c \xi_{\bf k} & \mathcal{D}_{\uparrow\uparrow} & \mathcal{D}_{\uparrow\downarrow} \\
-2\lambda^c \xi^\ast_{\bf k} & \bar{\varepsilon}^c_{\bf k}+\mu_BH &  \mathcal{D}_{\downarrow\uparrow} & \mathcal{D}_{\downarrow\downarrow} \\
\mathcal{D}^\ast_{\uparrow\uparrow} & \mathcal{D}^\ast_{\downarrow\uparrow} & \bar{\varepsilon}^f_{\bf k}-\mu_BH & -2\lambda^f\xi^{}_{\bf k} \\
\mathcal{D}^\ast_{\uparrow\downarrow} & \mathcal{D}^\ast_{\downarrow\downarrow} & -2\lambda^f\xi^\ast_{\bf k} & \bar{\varepsilon}^f_{\bf k}+\mu_BH
\end{pmatrix}
\end{align}
where $\xi_{\bf k}=\sin k_y+i\sin k_x$ and $\bar{\varepsilon}^{c/f}_\mathbf{k} = \varepsilon^{c/f}_\mathbf{k}+Un^{f/c}$. The Hamiltonian $\mathcal{H}_{\text{BdG}}(\mathbf{k})$ can be diagonalized numerically by using some standard routines and then the quasiparticle spectrum and eigenstates are delivered. In order to address explicitly the spin dependence on the EC order parameter as a result of electron-hole coherence, hereafter, we use $\Delta_{\sigma\sigma'}\equiv\mathcal{D}_{\sigma,-\sigma'}$. The EC order parameters $\Delta_{\sigma\sigma'}$ thus can be evaluated self-consistently. To characterize the topological properties in the electron-hole system, we evaluate here the Chern number $C$ by using the Fukui-Hatsugai-Suzuki (FHS) method~\cite{JPSJ.74.1674}, which discretizes the Brillouin zone and evaluates the Berry curvature via the gauge-invariant lattice formulation, one has
\begin{equation}
C = \frac{1}{2\pi i}\sum_{\mathbf{k}n} F^{[n]}(\mathbf{k}), 
\end{equation}
where $F^{[n]}(\mathbf{k}) = \ln[U^{[n]}_x(\mathbf{k}) U^{[n]}_y(\mathbf{k}+\hat{x}) U^{[n],-1}_x(\mathbf{k}+\hat{y}) U^{[n],-1}_y(\mathbf{k})]$ is the Berry curvature and $U^{[n]}_x(\mathbf{k})$ and $U^{[n]}_y(\mathbf{k})$ are link variables derived from the occupied eigenstates~\cite{JPSJ.74.1674,PRA.99.043611}. A nonzero Chern number indicates topological order reflecting the nontrivial topology induced by the effects of SOC and the magnetic field~\cite{PNAS.121.e2401644121}.

{\it Numerical results.}---
In order to inspect the topological properties of the ECs in the electron-hole system, we discuss here the numerical results addressing the EC order parameters $\Delta_{\sigma\sigma'}$ and Chern number $C$ evaluated for the 2D square-lattice with $N=500\times 500$ sites at zero temperature. Without loss of the generality, $t^c=1$ is set as the unit of energy and $t^f=-1$ \nham{for a nearly mass-balanced situation typically specified in TMDs~\cite{PRB.86.241401,PRB.88.085440}. Stimulated from the 2$\%$ strained 1$T$-WSSe material~\cite{PRB.103.195114}, we fix the SOC strengths of conduction and valence electrons to be $\lambda^c=0.2$ and $\lambda^f=0.5$, respectively~\cite{note1}.} Assuming that the system settles in the SM state before turning the interaction and magnetic field, we choose $E^c-E^f=1$. The chemical potential $\mu$ is then varied to fix the total electronic density $n^c+n^f=2$ as in the case of half-filling band. 

\begin{figure}[h]
\includegraphics[width=0.45\textwidth]{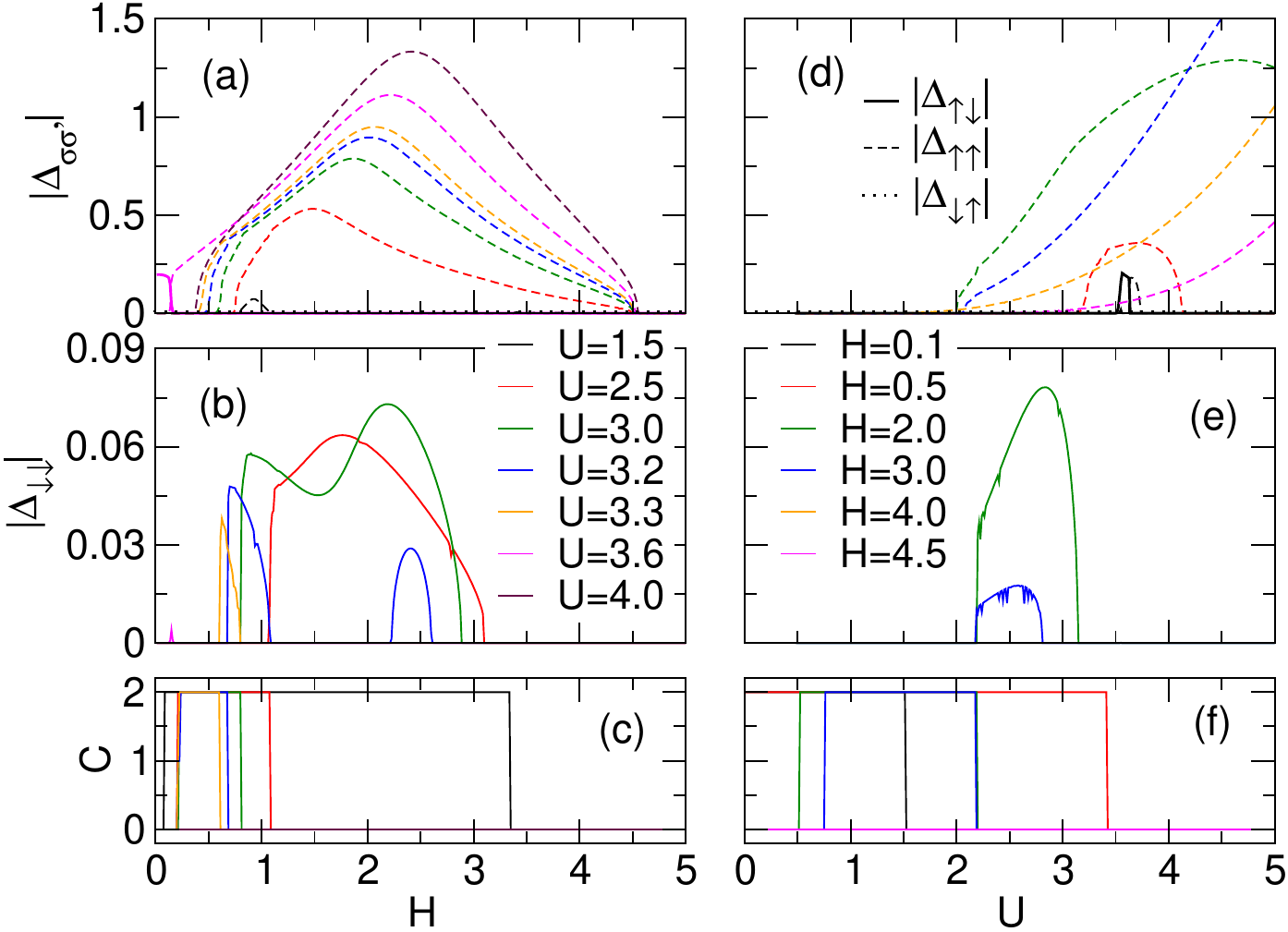}
\caption{Magnitude of EC order parameters $|\Delta_{\sigma\sigma'}|$ (first two rows) and Chern number $C$ (last row) as functions of magnetic field $H$ at different values of Coulomb interaction $U$ (left column) and as functions of $U$ at some values of $H$ (right column).}
\label{f1}
\end{figure}

Firstly, we analyze the behavior of the magnitude of EC order parameters $|\Delta_{\sigma\sigma'}|$ and Chern number $C$ under the influence of magnetic field $H$ and Coulomb interaction $U$. Figure~\ref{f1} presents $|\Delta_{\sigma\sigma'}|$ and $C$ as functions of $H$ at fixed $U$ (left column) and as functions of $U$ at fixed $H$ (right column). Figures~\ref{f1}(a) and~\ref{f1}(d) show that $|\Delta_{\sigma\sigma'}|$ emerges only when $U$ exceeds a critical threshold and $H$ is sufficiently low, reflecting the necessity of strong local electron-hole pairing to overcome the kinetic energy and minimal Zeeman splitting, stabilizing excitonic coherence marked by the opening of an excitonic gap~\cite{EM04,NC.5.4555,LGSXLLCM14}. At intermediate $U$ ($U\sim 3.6$), a singlet EC ($|\Delta_{\uparrow\downarrow}|\neq 0$) appears at low $H$, driven by isotropic $s$-wave pairing of opposite spins, which transits to a spin-up triplet EC ($|\Delta_{\uparrow\uparrow}| \neq 0$) as $H$ increases. This singlet-to-triplet transition, induced by growing Zeeman splitting, favors $\uparrow$-$\uparrow$ pairing by lowering the energy of aligned spins, with SOC enhancing the triplet’s stability~\cite{NN.9.611,PRB.100.041402}. At large $U$, the singlet EC at low $H$ is suppressed due to the Hartree shift, which increases the effective band gap, driving the system toward a normal SC state. However, increasing $H$ in this regime promotes the spin-up triplet EC by amplifying Zeeman splitting. In a narrow region of intermediate $H$ and $U$, one also finds a coexistence of a spin-down triplet EC ($|\Delta_{\downarrow\downarrow}| \neq 0$) with the dominant spin-up triplet EC, reflecting spin competition as Zeeman splitting slightly favors $\downarrow$-$\downarrow$ pairs [see Fig.~\ref{f1}(b) and~\ref{f1}(e)]. A nonzero Chern number ($C = 2$) is found in the spin-up triplet EC at intermediate $H$ and $U$, away from the singlet and spin-down triplet ECs [see Fig.~\ref{f1}(c) and~\ref{f1}(f)], indicating a nontrivial topological state. This topological order, driven by SOC-enhanced Berry curvature, highlights a novel topological spin-up triplet EC phase in the presence of the SOC and external magnetic field.

\begin{figure}[h]
\includegraphics[width=0.43\textwidth]{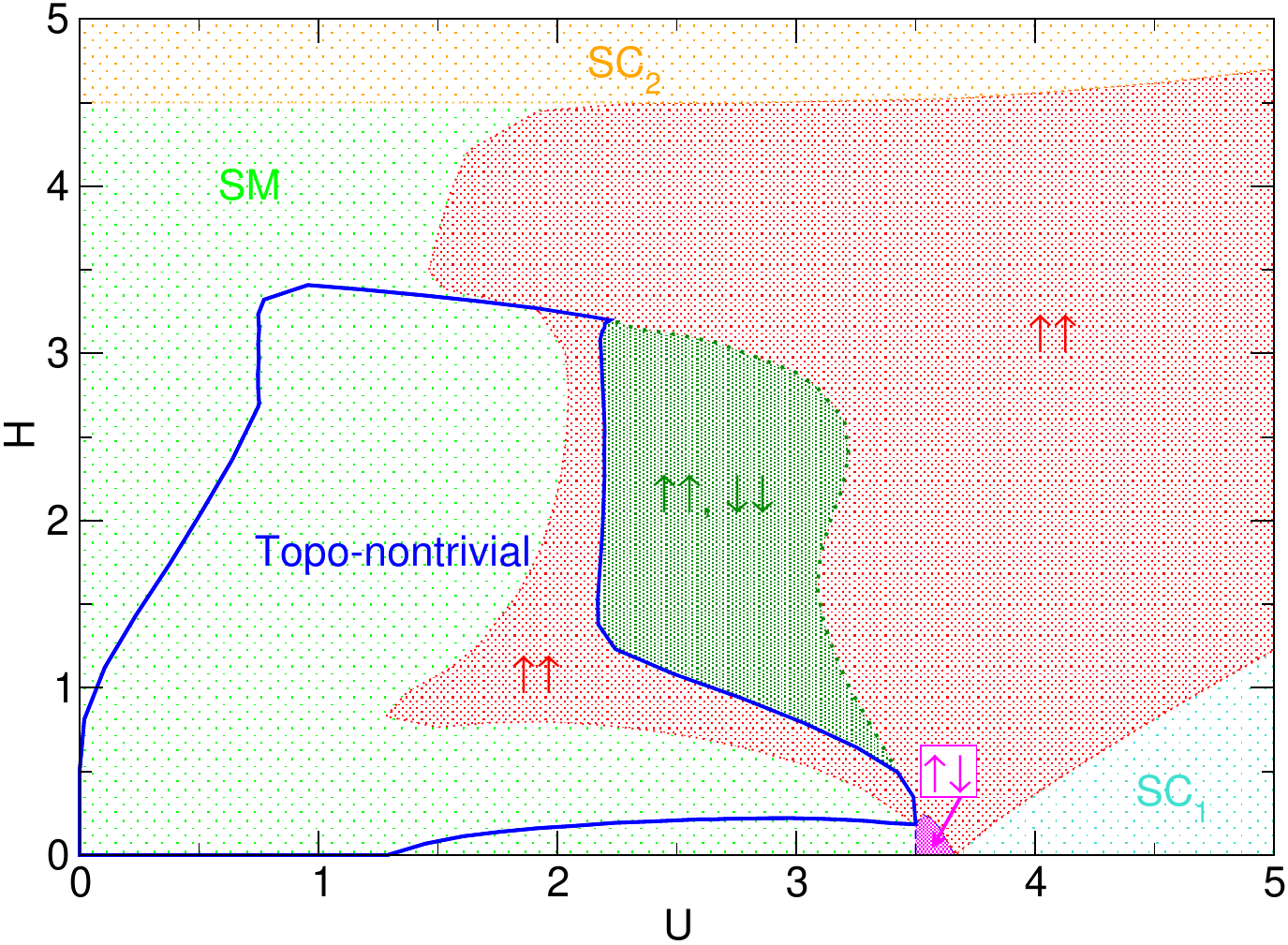}
\caption{Ground state phase diagram of the 2D electron-hole system in the $H-U$ plane. EC can be stabilized either in singlet (magenta), spin-up triplet (red), or spin-up and spin-down triplet coexistence (green). The SM state is marked in light-green, and the SC state is marked either in light-blue (SC$_1$) or in light-orange (SC$_2$). The topological nontrivial region is bounded by the blue solid line.}
\label{f2}
\end{figure}

Figure~\ref{f2} maps the rich ground state phases in the $H-U$ diagram, revealing a novel feature where the spin-up triplet EC exhibits topological order uniquely outside of the region where the spin-down triplet EC emerges. At weak interaction $U$ and small magnetic field $H$, the system remains in the SM phase characterized by partially overlapping electron and hole bands with an inverted spin-orbit structure. In this regime, the system exhibits a nontrivial topology inherited from the SOC-driven band inversion, analogous to earlier studies on topological SMs~\cite{PRB.87.155415,PRB.107.115147,RMP.93.025002}. The SM region extends up to a critical value of Coulomb interaction, beyond which the electron-hole attraction triggers ECs. Increased $U$ and moderate $H$ favor a spin-up polarized triplet EC state, where both the electron and hole reside in the spin-up channel and establish the excitonic coherence. Importantly, because condensation occurs within a single spin sector without mixing the conduction and valence bands across spin channels, the band inversion remains intact and the nontrivial topology is preserved~\cite{NN.15.367}. Here, excitonic pairing opens a gap within the inverted bands, not removing their topological character. In the case of small $H$, one finds singlet EC dominates due to the isotropic pairing. In this phase, the excitonic order parameter mixes spin-up and spin-down sectors, inducing a gap that symmetrically opens across the previously inverted bands. This inter-spin-channel hybridization eliminates the spin-filtered band structure and the nontrivial topology is not further protected. Consequently, one finds a topologically trivial EC state, as similarly observed in theoretical treatments of spin-singlet excitonic systems where the band inversion is annihilated by spin mixing~\cite{JPCM.27.333201,PRL.103.026402}. The vanishing of the Chern number directly reflects the loss of the SOC-driven inverted character after the singlet condensation.

A more intricate phase is the coexistence of the spin-up and the spin-down triplet ECs emerging at intermediate $U$ and higher $H$. In this regime, the Zeeman field favors spin imbalance, yet the interaction is strong enough to induce pairing in both spin channels. However, because both spin bands develop gaps, the original band inversion is removed across both channels, leading to a trivial topological character. The coexistence of two triplet condensates effectively eliminates the spin-filtered structure~\cite{NN.15.367}. Once $U$ is sufficiently large but $H$ remains small, the system undergoes a transition into a SC$_1$ phase, where the Hartree shift pushes the bands apart, creating a trivial gap without ECs. The topology is destroyed because the Hartree field cancels the band inversion originally protected by SOC, leading to conventional SC behavior. A similar trivial SC phase (SC$_2$) appears at large $H$, where the Zeeman splitting prevails over the SOC, lifting the spin degeneracy and leading again to a fully gapped SC state. In the SC$_2$, strong spin polarization depletes one spin species, suppressing any possibility of coherent excitonic pairing, consistent with general expectations for large magnetic field-induced trivialization~\cite{PRB.87.155415}. The essential physical mechanism governing topological preservation is thus the selectivity of excitonic pairing. Only when condensation occurs within a single spin sector -- preserving the inverted band structure -- is the Chern number maintained. Any other phase that either mixes spin channels or gaps both spins trivially destroys the topology. This principle reflects the behavior in spin-filtered Chern insulators and spin-polarized quantum anomalous Hall systems, where single-spin channel protection is crucial~\cite{PRB.104.195419,RMP.82.1539,RMP.95.011002}. \nham{In TMD polymorphs, the local Coulomb interaction typically ranges from 0.9 to 2.4 eV~\cite{PRB.109.125108}, while the hopping amplitude is on the order of 0.2 eV~\cite{PRB.88.085440}. Within this realistic parameter regime, the topological spin-up triplet EC found at $U\sim 2-4$ and $H\geq 0.2$ in Fig.~\ref{f2} is therefore experimentally relevant~\cite{note2}. These findings provide a concrete theoretical framework for realizing and stabilizing topological triplet EC states in 2D TMDs under external tuning~\cite{NN.15.367,PRB.87.155415}.}

\begin{figure}[t]
\includegraphics[width=0.47\textwidth]{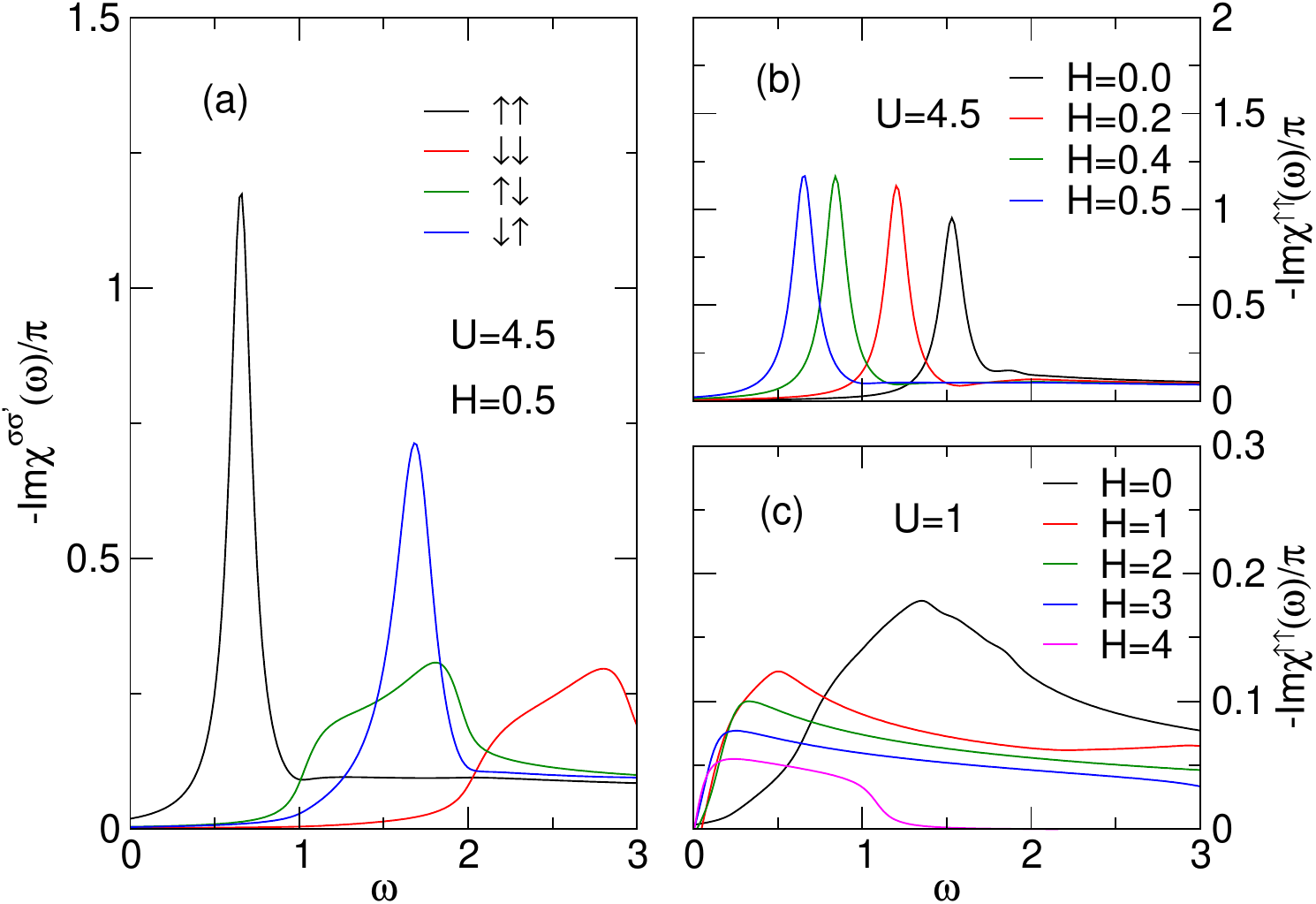}
\caption{The imaginary part of the dynamical excitonic susceptibility functions at zero momentum $\text{Im} \chi^{\sigma\sigma'}(\omega)$ for all spin configurations at $U=4.5$ and $H=0.5$ (a) and for spin-up triplet only with different values of $H$ at $U=4.5$ (b) and at $U=1$ (c).}
\label{f3}
\end{figure}

To further elucidate the nature of the excitonic fluctuations leading to the condensations, we analyze the imaginary part of the dynamical excitonic susceptibility function at zero-momentum, $\text{Im} \chi^{\sigma\sigma'}(\omega)$, evaluated by employing the RPA~\cite{PRB.106.035120,PRB.107.115106,PRB.109.085105,PRB.110.235143}. In Fig.~\ref{f3}, we address $\text{Im} \chi^{\sigma\sigma'}(\omega)$ for different Coulomb interactions and magnetic fields, illustrating the fluctuation signature that precedes the EC transitions observed in the phase diagram. At $U=4.5$ and $H=0.5$, although long-range excitonic order has not yet set in, the susceptibility already shows a strong spin selectivity that the spin-up triplet channel $\text{Im} \chi^{\uparrow\uparrow}(\omega)$ exhibits a sharp low-frequency peak, while the spin-down and spin-singlet channels are much weaker. This strong enhancement suggests a precursor fluctuations or `halo' state toward the spin-polarized triplet excitonic condensate due to the strong Coulomb interaction in the SC state [see Fig.~\ref{f3}(a)]. As increasing $H$ approaching the EC threshold, the low-energy peak of $\text{Im} \chi^{\uparrow\uparrow}(\omega)$ sharpens and shifts toward zero frequency, evidencing the progressive softening of the excitonic mode and signaling the proximity to the spin-up triplet EC [Fig.~\ref{f3}(b)]. Such spin-selective enhancement specifies the impact of the external fields promote spin-polarized excitonic correlations by splitting the conduction and valence bands asymmetrically~\cite{S.351.688,PRB.100.041402,Nat.567.66}. In the contrast, at a small interaction, the peaks become broader and one finds the Drude-type behavior at low-frequencies [see Fig.~\ref{f3}(c)] indicating the SM state~\cite{PRB.98.115139,AM76}. Notably that, for $H\leq 3$, corresponding to the topologically nontrivial SM region, $\text{Im} \chi^{\uparrow\uparrow}(\omega)$ retains a clear low-frequency enhancement, indicating that the underlying topological band structure actively promotes low-energy spin-up triplet excitonic fluctuations. When $H\sim 4$, however, the system transits into a trivial SM state and the low-energy spectral weight in $\text{Im} \chi^{\uparrow\uparrow}(\omega)$ diminishes. This supports the idea that the topological protection amplifies excitonic fluctuations at low energies in the SM state. Otherwise, the excitonic coherence is weakened unless further reinforced by strong interactions.

{\it Summary and conclusion.}---We have theoretically demonstrated the emergence of a topological spin-up triplet excitonic condensate in two-dimensional interacting electron-hole systems under Rashba spin-orbit coupling and external magnetic fields. Unlike conventional excitonic states, this condensate exhibits a nonzero Chern number, stabilized by spin-selective pairing that preserves the SOC-driven band inversion. The pronounced spin-polarized excitonic fluctuations preceding the condensate are also revealed in the signatures of the dynamical excitonic susceptibility. These findings establish an alternative mechanism for topological excitonic condensation stability and suggest a \nham{monolayer of transition metal dichalcogenides such as distorted 1$T$-phase Janus WSSe or some twisted van der Waals heterostructures} as promising platforms for detecting it in real systems.


This research is funded by the Vietnam National Foundation for Science and Technology Development (NAFOSTED) under Grant No. 103.01-2023.43.


\begin{thebibliography}{86}
\bibitem [{not({\natexlab{a}})}]{note1}%
  \BibitemOpen
  \href@noop {} \bibinfo {note} {Indeed, 2$\%$ strained
  1$T$-WSSe exhibits Rashba SOC parameters $\alpha_v\approx 0.5$ eV\r{A} in the
  valence band and $\alpha_c\approx 0.2$ eV\r{A} in the conduction
  band~\cite{PRB.103.195114}. Given typical lattice constants $\sim 5$
  \r{A}~\cite{PRB.103.195114} and hopping magnitudes $\sim 0.2$
  eV~\cite{PRB.88.085440}, these translate into effective dimensionless SOC
  strengths $\lambda_f=0.5$ and $\lambda_c = 0.2$ in our normalized units
  (lattice constant $a = 1$).}\BibitemShut {Stop}%
\end{thebibliography}

%

\end{document}